\documentclass[twocolumn,fleqn,10pt]{wlscirep}
\usepackage{graphicx,hyperref,amsmath,amsfonts}
\usepackage{epstopdf,color,bm,multirow,rotating,soul,comment}
\usepackage{parskip}

\title{Multiresonator quantum memory-interface}
\author[1,2,*]{S.A. Moiseev}
\author[1,2]{K.I. Gerasimov}
\author[3]{R.R. Latypov}
\author[1,2]{N.S. Perminov}
\author[3]{K.V. Petrovnin}
\author[3]{O.N. Sherstyukov}
\affil[1]{Kazan Quantum Center, Kazan National Research Technical University n.a. A.N.Tupolev-KAI, 10 K. Marx, Kazan 420111, Russia}
\affil[2]{Zavoisky Physical-Technical Institute of the Russian Academy of Sciences, 10/7 Sibirsky Tract,
Kazan 420029, Russia}
\affil[3]{Kazan Federal University, 18 Kremlyovskaya Str., Kazan 420008, Russia}
\affil[*]{s.a.moiseev@kazanqc.org}

\begin{abstract}
\textbf{
In this paper we experimentally demonstrated a broadband microwave scheme suitable for the multiresonator quantum memory-interface. The microwave scheme consists of the system of composed mini-resonators strongly interacting with a common broadband resonator coupled with the external microwave waveguide. We have implemented the controllable tuning of the mini-resonator frequencies and coupling of the common resonator with the external waveguide for the implementation of the impedance matched quantum storage. The storage of microwave pulses with an efficiency of 16.3$\%$ has been shown experimentally at room temperature. The possible properties of the proposed scheme for mini-resonators with high-Q at low temperatures are discussed. The obtained results pave the way for the implementation of superefficient broadband microwave quantum memory-interface.
}
\end{abstract}

\begin{document}

\flushbottom
\maketitle
\thispagestyle{empty}

\noindent 
The development of the quantum memory (QM) as well as its effective light-media quantum interface are of decisive importance for quantum information technologies \cite{Devoret2013,Kurizki2015,Hammerer2010}. Impressive experimental results on the way to the effective optical QM were achieved in the last decade \cite{Hedges2010,Hosseini2011,Cho2016}. Recently the developed approaches stimulated active studies for the elaboration of the microwave QM which becomes highly important for the creation of a universal superconducting quantum computer \cite{Grezes2014,Gerasimov2014,Flurin2015,Pfaff2017}. The microwave QM should be able to store many short microwave pulses with the high efficiency \cite{Gambetta2017} and to satisfy very strong requirements of multiqubit quantum processing and error correction procedures \cite{Taminiau2014}. In the practical implementation of multiqubit QM, it is assumed to satisfy a quite strong coherent interaction of microwave qubits with many \cite{Hartmann2008,Roy2017} information carriers, in particular electron nuclear spins of NV-centers in diamond \cite{Jiang2009} and rare-earth ions in inorganic crystals \cite{Zhong2015}.

One of the promising approaches to constructing a QM is based on the spin/photon echo effect in resonant ensembles of electron or nuclear spins \cite{Hahn1950,Kurnit1964,Moiseev2001,Tittel2009}, where a strong coupling of microwave photons with quantum electrodynamics cavity mode also plays a crucial role for the effective reversible transfer of quantum information from the flying qubits to the long-lived spin coherence \cite{Afzelius2010,Moiseev2010,Sabooni2013_1,Grezes2014,Krimer2016,Arcangeli_2016}. It is possible to increase the quantum efficiency for the broadband interface in this approach via spreading the impedance matching condition to a wider range of working frequencies \cite{Moiseev2010}. However an effective solution of this problem remains unknown for high-Q resonator that strongly limits the spectral width of QM by the overly narrow resonator linewidth \cite{Flurin2015}.

In this work, starting from the AFC protocol of the photon echo QM \cite{Riedmatten2008} in the single mode cavity \cite{Afzelius2010}, we experimentally showed that the multiresonator (MR) quantum memory-interface (QMI) can be efficiently implemented for broadband microwave fields. To enhance dramatically the interaction with the signal microwave field, we used a set of a small number of mini-resonators coupled to a common broadband resonator that makes it possible to achieve a storage efficiency of 16.3$\%$ for short microwave pulses.

The constructed QMI prototype has promising technical properties: compactness, low cost and ease of fabrication, which are convenient for controlling the field dynamics and integration of the MR scheme into the microwave circuits of quantum processing. This technical solution allowed us to experimentally investigate the basic fundamental properties of the MR scheme, which can be implemented on another on-chip platform at low temperatures \cite{Brecht_2016,Zhong2017} with the considerably higher quantum efficiency.

\begin{figure*}[htb]
\begin{center}
\includegraphics[width=0.95\textwidth]{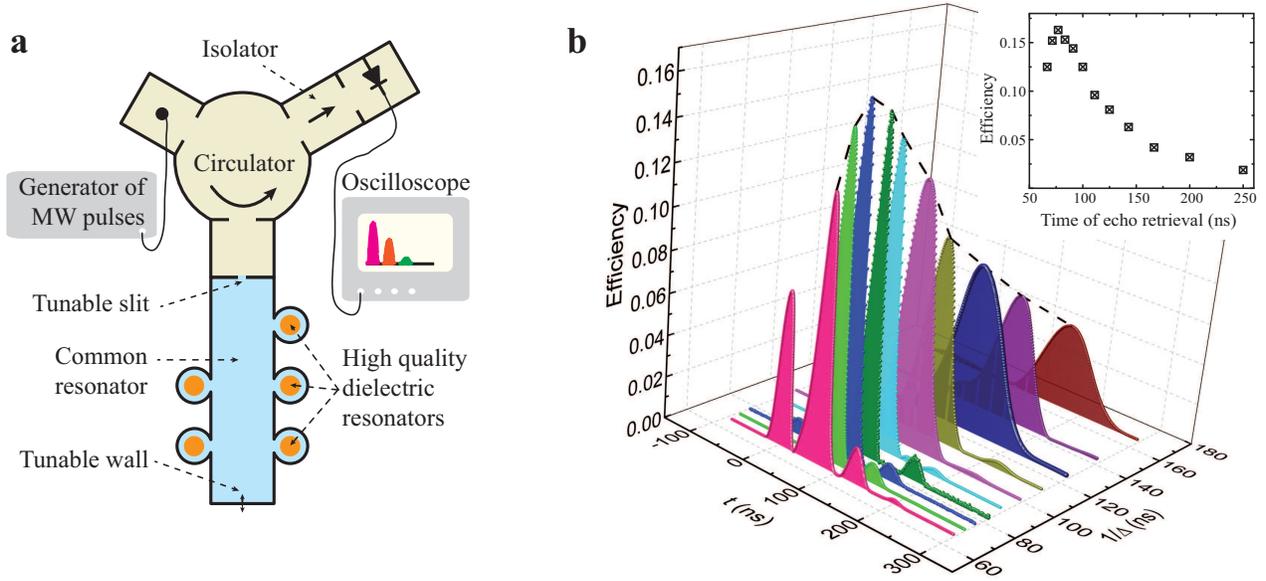}
\caption{\textbf{a}: Experimental setup and microwave prototype of the MR QMI scheme (blue color of filling): the large common resonator is connected with 5 cylindrical brass mini-resonators, the frequencies of which can be tuned by using the handles fixing the lengths of the cylinders. \textbf{b}: Reflected (t = 0) and echo signals from MR circuit at the varied spectral detuning $\Delta$ between mini-resonators. Dashed envelope and black square dots in the insert of figure \textbf{b} show the dependence of the efficiency on the storage time. Signals were normalized to the amplitude of input signals, which are not shown.}
\label{SetupAnd3D}
\end{center}
\end{figure*}

\textbf{QMI prototype}
\noindent
The scheme of the QMI prototype and experimental setup are shown in figure \ref{SetupAnd3D}\textbf{a}. The prototype contained five cylindrical brass resonators (mini-resonators) filled by the dielectric inserts (small cylindrical resonators) characterized by the quite large dielectric permittivity $\epsilon\sim29$. The mini-resonators had a variable length to provide fine tuning of the resonant frequency. Each mini-resonator was coupled to the large common rectangle broadband resonator through a thin nontunable slit. The large resonator was made of a copper waveguide (23$\times$10mm). It had one tunable wall to adjust its resonator frequency to provide a strong coupling of the input field with the mini-resonator system. The input tunable slit to the common resonator can be partially or fully open. The last regime was used to adjust the frequencies of mini-resonators to the periodic spectral comb with equal spectral distances $\Delta$.

\begin{figure}[htb]
\includegraphics[width=0.49\textwidth]{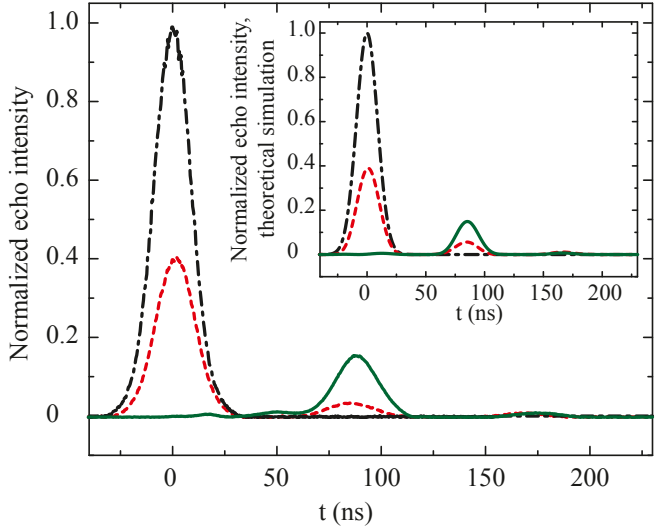}
\caption{Experimental curves of the normalized echo intensity for $\Delta=12$MHz: black dot-dashed line -- input pulse, red dashed line -- echo for open configuration (when the input slit is fully open), green solid line -- echo for optimal $\kappa$ (impedance matching condition). Theoretical curves of the normalized echo intensity for $\Delta=12$MHz are shown in the insert. Type and color of the lines for echo at the optimal $\kappa$ are the same as in the experimental curves in figure \ref{SetupAnd3D}\textbf{b}.}
\label{echo_sim_exp}
\end{figure}

\textbf{Echo and efficiency.}
\noindent
An Elexsys E580 ESR spectrometer (Bruker) at X-band (9.8GHz) was used in the preliminary echo experiments, where the efficiency of the retrieved echo was $\sim 4\%$ that corresponds to the theoretical red (dashed) curve in figure \ref{echo_sim_exp}. In these experiments, only broadband microwave pulses with the rectangular form can be used. Such pulses had no possibility for the effective implementation of the impedance matching condition and led to the very limited efficiency.

To improve the impedance matching condition, we used a home-built setup (figure \ref{SetupAnd3D}\textbf{a}). An Agilent E8267D vector signal generator with the amplitude modulation by an arbitrary waveform generator was used here as a source of microwave pulses with the Gaussian temporal form. The generator was coupled to the rectangular waveguide and through a standard X-band circulator to MR QMI. The reflected signal and emitted echo pulses were detected by the detector section. We used an isolator to avoid small undesirable signals reflected from the detector section and passed again to the generator.

We have performed a series of experiments with different spectral detunings $\Delta$ varied between 4 and 15MHz as shown in figure \ref{SetupAnd3D}\textbf{b}. Microwave classical Gaussian pulses with the spectral width up to 48MHz were stored in MR QMI and then were retrieved automatically by MR QMI in the time range of 66--250ns. As expected, the time of the echo emission was $1/\Delta$ and the efficiency strongly depended on the detuning $\Delta$. The dependence of the echo efficiency on the detuning $\Delta$ is in qualitative agreement with the analytical estimation of the quantum efficiency $\eta$ (\ref{eta_estimation}).

We found the analytical estimation of the first echo pulse retrieval by using equation (\ref{gen_eq}) (see section \textbf{Physical model}) at the optimal coupling $\kappa$:
\begin{align}\label{eta_estimation}
& \eta=\left[1+\frac{2\gamma_r\Delta}{g^2}\right]^{-2}\operatorname{exp}\left\{-\frac{2\gamma}{\Delta}\right\},
\end{align}
where we assumed the pulse spectrum $\delta\omega_f\sim(N-1)\pi\Delta$, $g=\left\langle g_n\right\rangle$ and  $\gamma=\left\langle \gamma_n\right\rangle$ are the average coupling constant and decay rate of mini-resonators, respectively, $\Delta=\left\langle\Delta_{n+1}-\Delta_n\right\rangle$ is the spectral distance between mini-resonators, $T_1=1/\Delta$ is the time of the signal recovery, the optimal coupling $\kappa =\kappa_{0}=2\gamma_r+g^2/\Delta$ of the common resonator with the external waveguide is determined by the impedance matching condition. For arbitrary $\kappa$ the estimation for efficiency is given by the formula $\eta=g^4/(\Delta^2\kappa^2)\operatorname{exp}(-2\gamma \ T_1)$.

The maximal achieved efficiency for the echo signal retrieval was 16.3$\%$ with the time delay $\sim77$ns ($\Delta=13$MHz). 
In this case the reflection of the input signal is highly suppressed due to the efficient impedance matching condition. 
One can see some partial reflection of the input pulse for other detunings $\Delta$ in figure \ref{SetupAnd3D}. 
We explain this fact by the complexity of the experimental setup leading to limited experimental capabilities for the perfect control of the impedance matching at different parameters of MR system. The reflection coefficient at the central frequency $\omega=0$ of MR system $R=[\kappa-2\gamma_r-g^2/\Delta]^2/[\kappa+2\gamma_r+g^2/\Delta]^2$.

The results of echo experiments for detuning $\Delta=12$ MHz are also shown in figure \ref{echo_sim_exp}, where the black (dot-dashed) curve is the input microwave pulse. The red (dashed) curve describes the reflected and two echo pulses when five coupled mini-resonators are connected to the external waveguide (the adjustable input slit of the large resonator was fully open). The green (solid) curve corresponds to the case when mini-resonators were included in the tunable large common resonator. Herein, the large resonator was coupled to the external waveguide where the coupling constant $\kappa$ was tuned to achieve the impedance matching condition. In this case, the efficiency of the echo emission also highly increased up to 15.4$\%$, while the second echo was almost absent as it is seen from the comparison of red (dashed) and green (solid) curves of figure \ref{echo_sim_exp} and the reflected signal at $t=0$ was also almost absent.

\textbf{Physical model.}
\noindent
The theoretical model of the considered memory corresponds to the so-called impedance matching QM on the photon echo in a single mode cavity \cite{Moiseev2010,Afzelius2010,Kalachev2013}, which was expanded further to the system of ring resonators connected with the nanooptical fiber \cite{ESMoiseev2016} and to other integrated schemes \cite{Yuan2016,Moiseev_2017_PRA}. Using the input-output formalism of quantum optics \cite{Walls} for the system studied in the work, we obtain the equations for field modes of mini-resonators $s_n(t)$ and of the common cavity field $a(t)$:
\begin{align}\label{gen_eq}
& \left[\partial_{t}+i2\pi\Delta_n+\gamma_n\right]s_n(t)+g_n^{*}a(t)=0,\\
& \nonumber \left[\partial_{t}+\frac{\kappa}{2}+i2\pi\Delta_r+\gamma_r\right]a(t)-\sum_ng_ns_n(t)=\sqrt{\kappa}a_{in}(t),
\end{align}
where $a_{in}(t)=[2\pi]^{-1/2}\int~d\omega e^{-i\omega t}f_\omega$ is the input pulse, $f_\omega$ is the spectral profile of the input pulse, for which the normalization condition is fulfilled $\int d\omega |f_\omega|^2=1$ for the single photon field, $\omega$ is the circular frequency counted from the central circular frequency of the radiation $\omega_0$, $\nu=\omega/(2\pi)$ is the conventional frequency, $\Delta_n$ are frequency detuning of mini-resonators, $n\in\{1,...,N\}$, $\gamma_n$ is the decay constant of the field mode in $n$-th mini-resonator, $\kappa$ is the coupling coefficient of the broadband resonator with the external waveguide, $g_n$ is the coupling constant of the common resonator mode with the field mode of the $n$-th mini-resonator, $\Delta_r$ is the frequency shift of the broadband resonator mode and $\gamma_r$ is the decay constant of this mode. We also ignored the Langevin forces \cite{Scully1997} in equations (\ref{gen_eq}) focusing on the study of the efficiency.

\textbf{Summary and outlook.}
\noindent
The implemented MR system made possible the independent experimental control of the optimal spectral detuning $\Delta_n$, coupling constants $g_n$ of mini-resonators and coupling constant $\kappa$ with the external waveguide that provided the impedance matching condition for the storage of microwave pulses as it is seen in figure \ref{echo_sim_exp}. Experimental results obtained under these conditions exhibit the quantum efficiency of 16.3$\%$ which is the highest for the storage of broadband microwave pulses. Namely, the spectrum width of the input pulse is much larger compared to the linewidth of each mini-resonator field mode.

The possibility of using a small number of microwave mini-resonators for efficient transfer of the broadband pulse from the common resonator to the spatially redistributed mini-resonators is a non-trivial problem associated with the increase in the storage time of the signal and the implementation of the optimal topology for the connection of all the mini-resonators in a spatially small common waveguide. The problem is caused by the fact that the field modes of the mini-resonators begin to strongly interact with each other in the common resonator and change strongly their initial frequencies. These changes should be accurately reflected in the initial frequencies to provide final frequencies be periodic as it is required \cite{Riedmatten2008} for the perfect QM. The super-high control of optimal spectral properties provides a superefficient quantum storage that is possible on the basis of the spectral-topological approach to MR QM \cite{Moiseev_2017_PRA,Perminov2017}.

Coming back to the presented microwave experiments, we note that the decay constants $\gamma_n$ and $\gamma_r$ could be much smaller in comparison with the coupling constants $g_n$ and spectral detuning $\Delta$ at helium temperature \cite{Megrant2012}. In this case, the realistic estimation of quantum efficiency (\ref{eta_estimation}) of our system at such temperature gives $\eta\cong1-4\gamma_r\Delta/g^2-2\gamma/\Delta\cong0.999$ for $\gamma\sim\gamma_r\sim 10^{-3}$MHz and $\Delta\sim g\sim4$MHz, which opens the way for the construction a highly efficient microwave QMI by using current technologies \cite{Kobe2017,Toth2017}.

\textbf{Acknowledgements.}
The research has been supported by the the Russian Science Foundation through the Grant No. 14-12-01333-P and by Russian Government Program of Competitive Growth of Kazan Federal University.
S.A.M. thanks A.V. Ustinov and M. Fleischhauer for useful discussions, K.I.S. thanks R.B. Zaripov for assisting in preliminary experiments and
K.V.P. thanks R.S. Kirillov for help in preparing the experiment.

\bibliographystyle{naturemag-doi}
\bibliography{MRQMI}

\end{document}